\documentclass[%
reprint,
amsmath,amssymb,
aps, 
prl
]{revtex4-2}

\usepackage{amsmath,amssymb}
\usepackage{graphicx}
\usepackage{color}
\usepackage{bm}
\usepackage{hyperref}
\usepackage{physics}
\usepackage{color}
\usepackage{soul}

\newcommand{\ee}{{\mathrm{e}}}
\newcommand{\ri}{{\mathrm{i}}}

\begin{document}

\title{Mutual Friction in Dissipative Gross-Pitaevskii Thermal Counterflow Turbulence}

\author{Kyo Yoshida}
\affiliation{%
Department of Physics, Institute of Pure and Applied Sciences, University of Tsukuba, 1-1-1 Tennoudai, Tsukuba 305-8571, Ibaraki, Japan}
\author{Hideaki Miura}
\affiliation{%
Department of Helical Plasma Research, National Institute for Fusion Science, 322-6 Oroshi-cho, Toki 509-5292, Gifu, Japan}
\author{Yoshiyuki Tsuji}
\affiliation{%
Department of Energy Science and Engineering, Nagoya University, Furo-cho, Chikusa, Nagoya 464-8601, Aichi, Japan}

\date{\today}

\begin{abstract}
We report numerical simulations of the dissipative Gross-Pitaevskii equation for a bulk region of thermal-counterflow turbulence. Quasistationary states are obtained over a range of forcing, damping, and healing-length parameters. The mutual-friction acceleration exhibits cubic scaling with the mean relative velocity between the superfluid and normal-fluid components, and the coefficient of this scaling is linked to the phenomenological damping parameter. The intervortex spacing follows the expected dimensional scaling in the weak-forcing regime. Comparison with a straight-vortex-line model suggests that the vortex-line orientations are nearly isotropic.
\end{abstract}

\maketitle

\paragraph{Introduction.---}
Quantum fluids, such as superfluids and Bose-Einstein condensates, exhibit fluid behavior while retaining some quantum properties. Unlike in classical fluids, vorticity in quantum fluids is concentrated on one-dimensional defects called vortex lines, and the circulation around each vortex line is quantized as $\kappa:=2\pi\hbar/m$, where $\hbar$ is the reduced Planck constant and $m$ is the particle mass. Despite this peculiarity, flows in quantum fluids can develop into turbulence just as flows in classical fluids do. In quantum fluid turbulence, the quantized vortex lines form a tangled structure. Considerable effort has been devoted to clarifying the similarities and differences between classical and quantum turbulence (see, e.g., Ref.~\cite{TsubotaKasamatsu2025}).

A prototypical example is thermal counterflow in superfluid $^4$He. In this flow, the superfluid and normal-fluid components move, on average, parallel and antiparallel to a large-scale temperature gradient, respectively. Mutual friction between the two components arises in the presence of quantized vortex lines. Experimental studies of thermal counterflow may be found, for example, in Refs.~\cite{Tough1982,Guoetal2010}; see Ref.~\cite{Sergeev2023} for a review of mutual friction.

The state of a superfluid or Bose-Einstein condensate is described by the complex order-parameter field $\psi(\bm{x},t)$, whose dynamics is governed by the Gross-Pitaevskii equation (GPE) \cite{PitaevskiiStringari2003}. Quantized vortex lines lie where $|\psi|=0$. When $|\psi|$ is nearly constant except in the immediate vicinity of the vortex lines, the vortex filament model (VFM), which treats only the vortex-line configuration dynamically, provides an efficient description. Schwarz \cite{Schwarz1985,Schwarz1988} incorporated mutual friction into the VFM under the assumptions of a spatially uniform normal-fluid flow and a fixed mean relative velocity between the superfluid and normal-fluid components. Simulations of the VFM were performed under periodic boundary conditions in all three spatial directions, corresponding to a bulk region far from the boundaries of a thermal-counterflow system. The VFM has since been extended to include coupling to nonuniform and dynamically evolving normal-fluid flows \cite{Kivotides2007,KhomenkoMishraPomyalov2017,YuiTsubotaKobayashi2018}.

The strength of the VFM is that it can treat systems close to laboratory scales. Its limitation is that it cannot resolve the vortex-core structure or compressible excitations such as density waves because it assumes an incompressible quantum fluid. In particular, vortex reconnections must be introduced artificially. These limitations can be overcome by returning to the GPE and incorporating the normal-fluid effect phenomenologically through a damping term in the dissipative GPE (DGPE) \cite{ChoiMorganBurnett1998}. The numerical cost is high because the healing length $\xi$, which is about $0.1\ \mathrm{nm}$ in liquid $^4$He, must be resolved. Relations between the damping parameter in the GPE and the friction coefficients in the VFM have been studied numerically for vortices with simple geometries; see, e.g., Refs.~\cite{KobayashiTsubota2006,MadarassyBarenghi2008,Wittmeretal2021}. DGPE simulations of vortex tangles have also been carried out in the context of isotropic turbulence \cite{KobayashiTsubota2005JPSJ}.

In this Letter, we report DGPE simulations representing a bulk region of thermal-counterflow turbulence. To realize this setting, we introduce a transformation of the order-parameter field together with a boundary condition for the transformed field. The simulations use $1024^3$ grid points, with the aim of achieving partial quantitative relevance to laboratory-scale systems, especially for mutual friction. We performed simulations for various values of the parameter set $(\tilde f, \gamma, \xi)$, defined below, to establish the basic properties of this numerical framework.

\paragraph{Dissipative Gross-Pitaevskii equation.---}
The dissipative Gross-Pitaevskii equation (DGPE) for the order-parameter field $\psi(\bm{x},t)$ is
\begin{equation}
\pdv{\psi}{t}=-\frac{\ri+\gamma}{\hbar}
\qty[-\frac{\hbar^2}{2m}\nabla^2 -\mu+ g|\psi|^2] \psi 
-\frac{\ri}{\hbar}V \psi, 
\label{eq:DGPE}
\end{equation}
where $g$ is the coupling constant characterizing particle interactions, $V(\bm{x})$ is the potential, $\mu$ is the chemical potential, and $\gamma$ is the dimensionless phenomenological damping parameter. For simplicity, we assume a local interaction and identify the condensate and noncondensate components with the superfluid and normal-fluid components, respectively, unless otherwise stated. This is an oversimplification for liquid $^4$He, so caution is required when comparing the DGPE with experiments. The DGPE describes the superfluid dynamics in the frame in which the normal-fluid component is at rest. The action of the normal fluid on the superfluid is incorporated through the damping term, whereas the back-action is neglected. To model thermal counterflow, we set $V(\bm{x})=-mfx_1$, where $f$ is a uniform external acceleration applied to the superfluid in the $x_1$ direction.

The order-parameter field can be written as $\psi(\vb*{x},t)=\sqrt{n(\vb*{x},t)} \ee^{\ri \varphi(\vb*{x},t)}$, where $n(\vb*{x},t)$ is the condensate number density and $\vb*{v}(\vb*{x},t):=(\hbar/m)\vb*{\nabla}\varphi(\vb*{x},t)$ is the superfluid velocity field. The velocity field is irrotational except at vortex lines, where $n(\vb*{x},t)=0$ and $\varphi(\vb*{x},t)$ is undefined. The circulation around a path enclosing a vortex line is quantized in units of $\kappa:=2\pi\hbar/m$. The density recovers from $n=0$ at the vortex line to the equilibrium value $\bar{n}=\mu/g$ over the healing length $\xi:=\hbar/\sqrt{2m\mu}$.

\paragraph{Scaling laws.---}
Vortex lines moving relative to the normal fluid generate mutual friction. In an appropriate range of $f$, the vortex lines are expected to develop into a tangle, and the system reaches a quasistationary state in which the mutual-friction force balances the external force. In this regime, $f$ may be identified with the acceleration due to mutual friction, acting in the opposite direction. We assume that the dominant contribution to mutual friction is determined by the global structure of the vortex tangle rather than by the microscopic density profile around a vortex line or by other dynamics such as density waves. Under this assumption, $f$ and the relative velocity $w$ are related only through $\kappa$, and not through $\xi$ or any other quantity with dimensions of length or time. Dimensional analysis then gives
\begin{equation}
f=C(\gamma) \kappa^{-1} w^3, 
\label{eq:f=w^3}
\end{equation}
where $C(\gamma)$ is a dimensionless coefficient that may depend on $\gamma$.

The characteristic length scale of the vortex tangle is the mean intervortex spacing $l:=\lambda^{-1/2}$, where $\lambda$ is the vortex-line density. Assuming that $l$ is determined only by $\kappa$ and $f$, dimensional analysis gives
\begin{equation}
  l= C_l(\gamma) \kappa^{\frac{2}{3}}f^{-\frac{1}{3}}, 
\label{eq:l=f^(-1/3)}
\end{equation}
where $C_l(\gamma)$ is another dimensionless coefficient that may depend on $\gamma$. If $l \sim \xi$, the density structure near neighboring vortex lines is no longer negligible. If $l \sim L$, boundary effects become apparent. Therefore, Eqs.~(\ref{eq:f=w^3}) and (\ref{eq:l=f^(-1/3)}) are expected to hold in the range $\xi \ll l \ll L$, or equivalently $\kappa^2 L^{-3} \ll f \ll \kappa^2 \xi^{-3}$. These scaling laws are naturally consistent with earlier results \cite{Vinen1957,SwansonDonnelly1985,Schwarz1988} based mainly on the VFM, i.e., in the limit $\xi \to 0$. Within the DGPE framework, the dependence on the damping parameter $\gamma$ is contained in the coefficients $C(\gamma)$ and $C_l(\gamma)$, but their functional forms cannot be determined by dimensional analysis alone.

\paragraph{Numerical simulation settings.---}  
We introduce the transformed field $\phi(\bm{x},t):=\ee^{-\ri (m f/\hbar) x_1 t}\psi(\bm{x},t)$. The DGPE is thereby converted into a time-evolution equation for $\phi$ in which the position $\bm{x}$ no longer appears explicitly.
We may then impose periodic boundary conditions, $\phi(\bm{x}+L \bm{e}_j,t)=\phi(\bm{x},t)$, where $\bm{e}_j$ ($j=1,2,3$) are unit vectors along the coordinate axes. Such boundary conditions cannot be imposed directly on the original field $\psi$, because the potential $V$ in Eq.~(\ref{eq:DGPE}) is not periodic. Let $\phi_{\bm{k}}(t):=L^{-3}\int \dd[3]{\bm{x}}\ee^{-\ri \bm{k}\cdot\bm{x}} \phi(\bm{x},t)$,
with $\bm{k}=(k_1,k_2,k_3)$ and $k_j=(2\pi/L)n_j$ for $n_j\in \mathbb{Z}$, denote the Fourier transform of $\phi(\bm{x},t)$.
Introducing the normalized variables $\tilde t:=\mu t/\hbar$, $\tilde\phi:=\sqrt{g/\mu}\,\phi$, and $\tilde f:=(m \xi/\mu) f$, we obtain
\begin{align}
{\pdv{{\tilde\phi}_{\bm k}}{\tilde t}} ({\tilde t})
&=-\qty(\ri+\gamma)
\bigg[\qty(\xi^2 |\bm{k}(\tilde t)|^2 -1){\tilde\phi}_{\bm{k}}(\tilde t)
  \nonumber\\
&\quad+\sum_{\bm{p}\bm{q}\bm{r}} \delta_{\bm{k}+\bm{p}-\bm{q}-\bm{r}}
{\tilde\phi}_{\bm{p}}^*(\tilde t){\tilde\phi}_{\bm{q}}(\tilde t) \tilde\phi_{\bm{r}}(\tilde t)
\bigg], 
\label{eq:DGPphik}
\end{align}
where $k_j(\tilde t):=k_j+(\tilde f \tilde t/\xi)\delta_{j1}$ and $\delta_{\bm{k}}=1$ for $\bm{k}=\bm{0}$ and $\delta_{\bm{k}}=0$ otherwise. In numerically solving Eq.~(\ref{eq:DGPphik}), the methods and settings, except for the damping and potential forcing, are the same as in Refs.~\cite{YoshidaMiuraTsuji2019,YoshidaMiuraTsuji2023}. Let $k_{\max}$ be the maximum resolved wavenumber and $\Delta k$ the wavenumber spacing.  To ensure $|\bm{k}(\tilde t)|<k_{\max}+\Delta k$ for all $|\bm{k}|<k_{\max}$ throughout the simulated time interval,
the replacement ${{\tilde\phi}^{\mathrm{new}}}_{\bm{k}}(\tilde t)={{\tilde\phi}^{\mathrm{old}}}_{\bm{k}-\Delta k\bm{e}_1}(\tilde t)$ is made at times $\tilde t= \qty(n + 1/2)\xi\Delta k/\tilde f\, (n\in \mathbb{N})$.  We set ${\tilde\phi}^{\mathrm{new}}_{\bm{k}}(\tilde t)=0$ when $\bm{k}-\Delta k\bm{e}_1$ lies outside the simulated range. The time-dependent wavenumber vector is then modified as
$
k_j(\tilde t):=k_j+\Phi(\tilde f \tilde t/\xi\Delta k)\Delta k\delta_{j1}
$,
where $\Phi(x):=x - \lfloor x+1/2\rfloor$.

We performed DGPE simulations for various parameter sets $(\tilde f, \gamma, \xi)$. The number of grid points in each coordinate direction was fixed at $N=1024$, and the system size was set to $L=2\pi$, so that $\Delta k=1$ and $k_{\max}=256$. The values of $\xi$ are $0.032$, $0.016$, and $0.008$, corresponding to $k_{\max}\xi\simeq 8$, $4$, and $2$, respectively.  Note that $k_{\max}\xi \gg 1$ is the condition for sufficient resolution.  By contrast, large-scale structures are captured only up to length scales substantially smaller than $L=196\xi$, $393\xi$, and $785\xi$.
The values of $\gamma$ are $0.025$, $0.05$, and $0.1$. For each pair $(\gamma, \xi)$, $\tilde f$ was varied from $10^{-5}$ to $6.4 \times 10^{-4}$
in steps of factors of $2$. The time step was set to $\Delta \tilde t =0.005$ for $\xi=0.032$ and $\Delta\tilde t=0.01$ for $\xi=0.016$ and $0.008$.

The main quantity monitored in the simulations is the averaged wavenumber $\bar{k}(\tilde{t}):=\sum_{\bm k} k_1(\tilde{t}) |\tilde{\phi}_{\bm k}(\tilde{t})|^2$ in the forcing direction. It is related to the momentum density as $p=\bar{n}\hbar \bar{k}$.  The mean relative velocity of the superfluid component with respect to the normal-fluid component is estimated as $w=(\hbar/m) \bar{k}$. The initial state for the first parameter
set was generated as in Ref.~\cite{YoshidaMiuraTsuji2019}, and the final states of preceding simulations were used as initial states for subsequent parameter sets. For each parameter set, the simulation was continued until at least quasistationarity was reached for $w$, with termination times ranging from $\tilde t=800$ to $14400$.

\paragraph{Relative velocity.---}
\begin{figure}[t]
\includegraphics[width=0.47\textwidth]{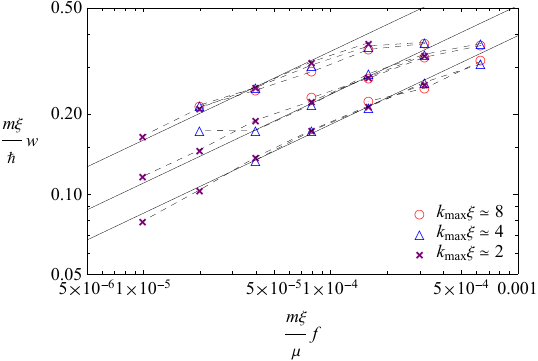}
\caption{Relative velocity $w$ versus acceleration $f$. Data from the same parameter set $(\gamma, \xi)$ are connected by dashed lines as a guide. Nearly overlapping dashed curves correspond to the same value of $\gamma$. The values of $\gamma$ are $0.025$, $0.05$, and $0.1$ in decreasing order of $w$. Solid lines show fits to the scaling law (\ref{eq:f=w^3}).}
\label{fig:veldif}
\end{figure}
The mean relative velocity $w$ in quasistationary states, averaged over suitable time intervals, is shown as a function of the acceleration $f$ for various parameter sets $(\gamma, \xi)$ in Fig.~\ref{fig:veldif}. For each fixed pair $(\gamma, \xi)$, the data approximately follow Eq.~(\ref{eq:f=w^3}). When $f$ is too large, $w$ tends to saturate and deviate from the scaling law. When $f$ is too small, $w$ becomes unstable, and the corresponding data are not shown. For fixed $(\tilde f,\gamma)$, the values of $w$ nearly collapse for different values of $k_{\max}\xi$. This suggests that $k_{\max}\xi=2$ already provides sufficient resolution for estimating bulk quantities such as $w$. Equation~(\ref{eq:f=w^3}) was fitted after excluding cases in which the forcing is too strong or too weak; the fits are shown by the solid lines. The fitted values are $C(0.025)=0.0078$, $C(0.05)=0.0024$, and $C(0.1)=0.052$, which suggest the form
\begin{equation}
C(\gamma)\simeq 1.8 \gamma^{\frac{3}{2}} .
\label{eq:Cgamma}
\end{equation}

\paragraph{Vortex-line configuration.---}
\begin{figure}
\centering
\includegraphics[width=0.35\textwidth]{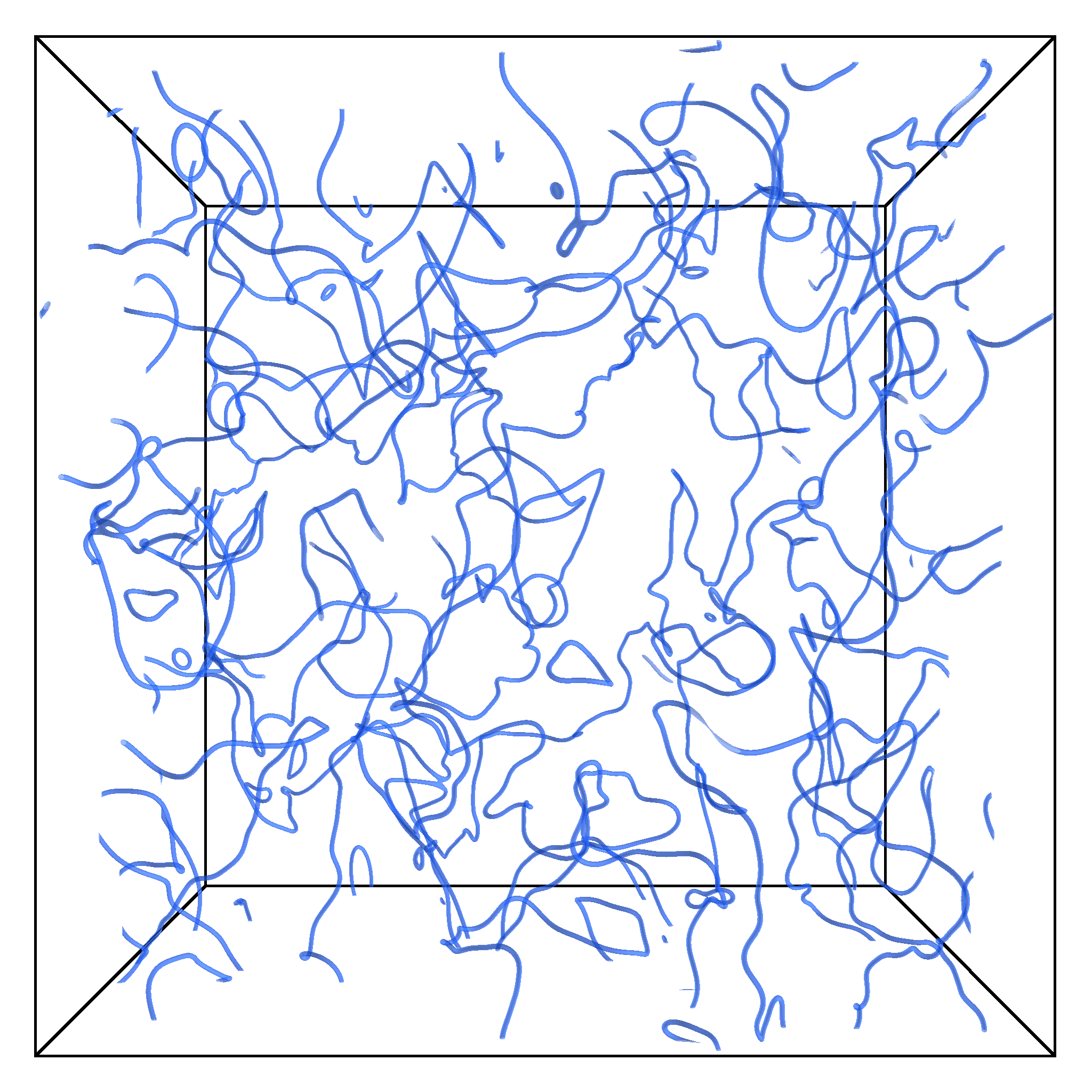}
\caption{Isosurface of $n=0.25\bar{n}$ in a quasistationary state for $(\tilde f, \gamma, \xi)=(0.8\times 10^{-4}, 0.025, 0.016)$. The external forcing points from left to right. The visualization data were produced using VISMO \cite{OhnoOhtani2014}.}
\label{fig:vortextangle}
\end{figure}
\begin{figure}
\includegraphics[width=0.45\textwidth]{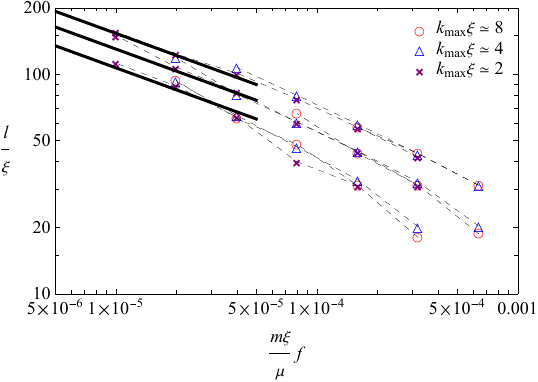}
\caption{Mean intervortex spacing $l$ versus acceleration $f$. Data from the same parameter set $(\gamma,\xi)$ are connected by dashed lines as a guide. Nearly overlapping dashed curves correspond to the same value of $\gamma$. The values of $\gamma$ are $0.025$, $0.05$, and $0.1$ in increasing order of $l$. Solid lines show fits to the scaling law (\ref{eq:l=f^(-1/3)}).}
\label{fig:lintv}
\end{figure}
As a representative example of the vortex tangle, Fig.~\ref{fig:vortextangle} shows the isosurface of $n=0.25\bar{n}$ in a quasistationary state.
The vortex lines are enclosed by these isosurfaces. Let $F({\tilde n})$ denote the volume fraction of the region in which the normalized density $n(\vb*{x})/\bar{n}$ is less than or equal to ${\tilde n}$. Then the intervortex spacing $l$ may be estimated from the relation $F({\tilde n})\simeq \pi (r({\tilde n}))^2/l^2$ for small $\tilde n$, where $r({\tilde n})$ is the distance from a stationary straight vortex line at which the normalized density equals ${\tilde n}$.
The mean intervortex spacing $l$ in quasistationary states, estimated from this relation with $\tilde n=0.25$ and $r(0.25)\simeq 0.952\xi$, is shown as a function of $f$ in Fig.~\ref{fig:lintv}. For each fixed pair $(\gamma,\xi)$, the data are connected by dashed lines as a guide. The slopes steepen as $f$ increases, making it difficult to identify a single exponent over the full range. The data approach Eq.~(\ref{eq:l=f^(-1/3)}) only in a small-acceleration range, $(m\xi/\mu)f < 5\times 10^{-5}$. The solid lines show fits to Eq.~(\ref{eq:l=f^(-1/3)}) in that range, giving $C_l(0.025)=2.3$, $C_l(0.05)=2.8$, and $C_l(0.1)=3.3$. The fitted values of $C_l(\gamma)$ increase slightly with $\gamma$.

\paragraph{Straight vortex-line model.---}
Consider a model vortex configuration in which the vortex lines are parallel to the $x_3$ axis, their intersection points with the $(x_1,x_2)$ plane are distributed nearly uniformly with mean spacing $l$, and they move in the $x_1$ direction with velocity $w$. The acceleration due to mutual friction acting on this configuration may be estimated from that acting on a cylindrical domain $r:=\sqrt{x_1^2+x_2^2} \le R$ of arbitrary height in the $x_3$ direction, where the moving vortex line lies on the $x_3$ axis at the instant considered and $\pi R^2=l^2$. The corresponding order-parameter field is
$\psi(\vb*{x})=\sqrt{n(r)}\ee^{\ri \theta + \ri(m/\hbar) w x_1}$,
where $\theta:=\arctan (x_2/x_1)$ and $n(r)$ is the density profile around a stationary straight vortex line. For a given order-parameter field $\psi(\vb*{x})$ and domain $D$, the frictional force acting on the superfluid in $D$ is given by the time derivative, due to the DGPE without the potential $V$, of the momentum density $\psi^*(-\ri \hbar \vb*{\nabla})\psi$ integrated over $D$. Using this estimate for the cylindrical domain and neglecting terms of order $w^2$, we obtain $ 
f_{\textrm{model}}=\gamma \qty(\mu w/\hbar)\qty(l/\xi)^{-2}\qty( 4\pi \ln \qty(l/\xi) -8.67 )
$
for large $l/\xi$ ($>16$), in the negative $x_1$ direction. In the derivation, we used the numerical solution for $n(r)$ in the range $r \le 16\xi$ and its asymptotic form at larger $r$.

\begin{figure}[t]
\includegraphics[width=0.45\textwidth]{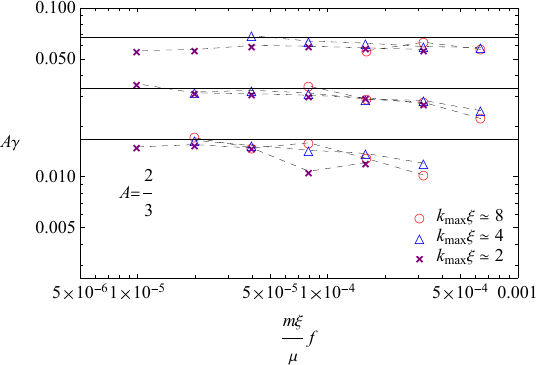}
\caption{Ratio $A$ of the observed acceleration $f$ to the model value $f_{\textrm{model}}$ in quasistationary states. Data from the same parameter set $(\gamma,\xi)$ are connected by dashed lines as a guide. The vertical axis is multiplied by $\gamma$ for visibility. Solid lines indicate $A=2/3$ for $\gamma=0.25$, $0.5$, and $1$ in increasing order.}
\label{fig:modelfit}
\end{figure}
Using the values of $w$ and $l$ measured in the quasistationary states, we evaluate the ratio $A:=f/f_{\textrm{model}}$, shown in Fig.~\ref{fig:modelfit}. In the model, all vortex lines are perpendicular to the velocity $w$. The acceleration is proportional to $\sin^2\vartheta$, where $\vartheta$ is the angle between the vortex line and the velocity, and $A=2/3$ for an isotropic distribution of vortex-line orientations. We find that $A\simeq 2/3$ or slightly smaller in most quasistationary states. This suggests that the vortex-line directions are distributed nearly isotropically. The slight decrease of $A$ with increasing $f$ may indicate a tendency for the vortex lines to align parallel or antiparallel to the velocity at large $f$. Other effects, however, may also influence $A$, including deformation of the density profile from $n(r)$, bending of the vortex lines, and nonuniformity of the vortex-line distribution.

\paragraph{Discussion.---}
Equation~(\ref{eq:Cgamma}) provides insight into the relation between the phenomenological damping parameter $\gamma$ and the coefficient $C$ characterizing thermal-counterflow turbulence. Relations between $\gamma$ and the friction coefficients $\alpha$ and $\alpha'$ in the VFM have been studied in Refs.~\cite{KobayashiTsubota2006,MadarassyBarenghi2008,Wittmeretal2021}. Note that $\alpha$ and $\alpha'$ characterize the interaction between a single vortex line and the normal-fluid component, whereas $C$ reflects the properties of vortex tangles generated by counterflow. Moreover, the coefficient $C$ obtained from the DGPE naturally includes the effect of vortex reconnections, which in VFM studies were introduced only artificially \cite{Schwarz1985,Schwarz1988,Kivotides2007,KhomenkoMishraPomyalov2017,YuiTsubotaKobayashi2018}.

Because the material-specific parameter $\xi$ appears explicitly in the DGPE and $C$ can be estimated from measurable quantities, namely the acceleration $f$ and the relative velocity $w$, the present study opens a route toward quantitative comparison between simulations and experiments on thermal-counterflow turbulence. Using $\mu\simeq 10^{-23}\ \mathrm{J}$ and $\xi\simeq 0.1\ \mathrm{nm}$ for liquid $^4$He, the corresponding simulated values are $f\sim 10^{9}-10^{10}\ \mathrm{m}/\mathrm{s}^2$ and $w\sim 10\ \mathrm{m}/\mathrm{s}$, substantially larger than the experimental values $f\sim 10-10^2\ \mathrm{m}/\mathrm{s}^2$ and $w\sim 10^{-1}\ \mathrm{m}/\mathrm{s}$ summarized in Fig.~24 of Ref.~\cite{Schwarz1988}. It is therefore not obvious that Eq.~(\ref{eq:f=w^3}), together with Eq.~(\ref{eq:Cgamma}), can be extrapolated to the experimental parameter range. Nevertheless, applying the scaling to the experimental data in Fig.~24 of Ref.~\cite{Schwarz1988}, together with the superfluid-density data in Ref.~\cite{DonnellyBarenghi1998}, yields the temperature dependent estimates $\gamma(1.174\ \mathrm{K})\simeq 0.002$, $\gamma(1.563\ \mathrm{K})\simeq 0.01$, and $\gamma(1.990\ \mathrm{K})\simeq 0.04$. Because nonlocal particle interactions are strong in liquid $^4$He, the values of $\gamma$ estimated from Eq.~(\ref{eq:DGPE}), which assumes a local interaction, should be regarded as effective parameters. Even with that caveat, the order of magnitude $\gamma=O(10^{-2})$ is consistent with the theoretical estimate in Ref.~\cite{ChoiMorganBurnett1998}.

Two directions for future work appear particularly natural. One is a macroscopic approach in which the system size is increased to test the scaling law for the intervortex spacing in Eq.~({\ref{eq:l=f^(-1/3)}}), which could not be confirmed clearly in the present study.  Such results would enable quantitative comparison with those obtained from experimental visualization of quantized vortices, such as in Refs.~\cite{TangBaoGuo2021,ChenMaruyamaTsuji2022}.  The coupling between the GPE and the normal-fluid flow may also be extended to include back-action, as proposed, for example, in Ref.~\cite{Brachetetal2023}. The other is a microscopic approach in which the simple DGPE used here is replaced by more elaborate models (see, e.g., Refs.~\cite{GriffinNikuniZaremba2009,BerloffBrachetProukakis2014}).

This work was performed on the ``Plasma Simulator'' (NEC SX-Aurora TSUBASA) at NIFS, with support from and under the auspices of the NIFS Collaboration Research program (NIFS22KISS017).


\begin{thebibliography}{24}%
\makeatletter
\providecommand \@ifxundefined [1]{%
 \@ifx{#1\undefined}
}%
\providecommand \@ifnum [1]{%
 \ifnum #1\expandafter \@firstoftwo
 \else \expandafter \@secondoftwo
 \fi
}%
\providecommand \@ifx [1]{%
 \ifx #1\expandafter \@firstoftwo
 \else \expandafter \@secondoftwo
 \fi
}%
\providecommand \natexlab [1]{#1}%
\providecommand \enquote  [1]{``#1''}%
\providecommand \bibnamefont  [1]{#1}%
\providecommand \bibfnamefont [1]{#1}%
\providecommand \citenamefont [1]{#1}%
\providecommand \href@noop [0]{\@secondoftwo}%
\providecommand \href [0]{\begingroup \@sanitize@url \@href}%
\providecommand \@href[1]{\@@startlink{#1}\@@href}%
\providecommand \@@href[1]{\endgroup#1\@@endlink}%
\providecommand \@sanitize@url [0]{\catcode `\\12\catcode `\$12\catcode
  `\&12\catcode `\#12\catcode `\^12\catcode `\_12\catcode `\%12\relax}%
\providecommand \@@startlink[1]{}%
\providecommand \@@endlink[0]{}%
\providecommand \url  [0]{\begingroup\@sanitize@url \@url }%
\providecommand \@url [1]{\endgroup\@href {#1}{\urlprefix }}%
\providecommand \urlprefix  [0]{URL }%
\providecommand \Eprint [0]{\href }%
\providecommand \doibase [0]{https://doi.org/}%
\providecommand \selectlanguage [0]{\@gobble}%
\providecommand \bibinfo  [0]{\@secondoftwo}%
\providecommand \bibfield  [0]{\@secondoftwo}%
\providecommand \translation [1]{[#1]}%
\providecommand \BibitemOpen [0]{}%
\providecommand \bibitemStop [0]{}%
\providecommand \bibitemNoStop [0]{.\EOS\space}%
\providecommand \EOS [0]{\spacefactor3000\relax}%
\providecommand \BibitemShut  [1]{\csname bibitem#1\endcsname}%
\let\auto@bib@innerbib\@empty
\bibitem [{\citenamefont {Tsubota}\ and\ \citenamefont
  {Kasamatsu}(2025)}]{TsubotaKasamatsu2025}%
  \BibitemOpen
  \bibfield  {author} {\bibinfo {author} {\bibfnamefont {M.}~\bibnamefont
  {Tsubota}}\ and\ \bibinfo {author} {\bibfnamefont {K.}~\bibnamefont
  {Kasamatsu}},\ }\href@noop {} {\emph {\bibinfo {title} {Quantum Hydrodynamics
  and Turbulence}}}\ (\bibinfo  {publisher} {Oxford University Press},\
  \bibinfo {year} {2025})\BibitemShut {NoStop}%
\bibitem [{\citenamefont {Tough}(1982)}]{Tough1982}%
  \BibitemOpen
  \bibfield  {author} {\bibinfo {author} {\bibfnamefont {J.~T.}\ \bibnamefont
  {Tough}},\ }\bibinfo {title} {Supefluid turbulence},\ in\ \href@noop {}
  {\emph {\bibinfo {booktitle} {Prog. Low Temp. Phys.}}},\
  Vol.~\bibinfo {volume} {8},\ \bibinfo {editor} {edited by\ \bibinfo {editor}
  {\bibfnamefont {D.}~\bibnamefont {Brewer}}}\ (\bibinfo  {publisher}
  {North-Holland Publishing Company, Amsterdam, New York, Oxford},\ \bibinfo
  {year} {1982})\ Chap.~\bibinfo {chapter} {3}\BibitemShut {NoStop}%
\bibitem [{\citenamefont {Guo}\ \emph {et~al.}(2010)\citenamefont {Guo},
  \citenamefont {Cahn}, \citenamefont {Nikkel}, \citenamefont {Vinen},\ and\
  \citenamefont {McKinsey}}]{Guoetal2010}%
  \BibitemOpen
  \bibfield  {author} {\bibinfo {author} {\bibfnamefont {W.}~\bibnamefont
  {Guo}}, \bibinfo {author} {\bibfnamefont {S.~B.}\ \bibnamefont {Cahn}},
  \bibinfo {author} {\bibfnamefont {J.~A.}\ \bibnamefont {Nikkel}}, \bibinfo
  {author} {\bibfnamefont {W.~F.}\ \bibnamefont {Vinen}},\ and\ \bibinfo
  {author} {\bibfnamefont {D.~N.}\ \bibnamefont {McKinsey}},\ }\href@noop {}
  {\bibfield  {journal} {\bibinfo  {journal} {Physical Review Letters}\
  }\textbf {\bibinfo {volume} {105}},\ \bibinfo {pages} {045301} (\bibinfo
  {year} {2010})}\BibitemShut {NoStop}%
\bibitem [{\citenamefont {Sergeev}(2023)}]{Sergeev2023}%
  \BibitemOpen
  \bibfield  {author} {\bibinfo {author} {\bibfnamefont {Y.~A.}\ \bibnamefont
  {Sergeev}},\ }\href@noop {} {\bibfield  {journal} {\bibinfo  {journal}
  {J. Low Temp. Phys.}\ }\textbf {\bibinfo {volume} {212}},\
  \bibinfo {pages} {251} (\bibinfo {year} {2023})}\BibitemShut {NoStop}%
\bibitem [{\citenamefont {Pitaevskii}\ and\ \citenamefont
  {Stringari}(2003)}]{PitaevskiiStringari2003}%
  \BibitemOpen
  \bibfield  {author} {\bibinfo {author} {\bibfnamefont {L.}~\bibnamefont
  {Pitaevskii}}\ and\ \bibinfo {author} {\bibfnamefont {S.}~\bibnamefont
  {Stringari}},\ }\href@noop {} {\emph {\bibinfo {title} {{Bose-Einstein}
  condensation}}}\ (\bibinfo  {publisher} {Oxford University Press},\ \bibinfo
  {year} {2003})\BibitemShut {NoStop}%
\bibitem [{\citenamefont {Schwarz}(1985)}]{Schwarz1985}%
  \BibitemOpen
  \bibfield  {author} {\bibinfo {author} {\bibfnamefont {K.}~\bibnamefont
  {Schwarz}},\ }\href@noop {} {\bibfield  {journal} {\bibinfo  {journal}
  {Physical Review B}\ }\textbf {\bibinfo {volume} {31}},\ \bibinfo {pages}
  {5782} (\bibinfo {year} {1985})}\BibitemShut {NoStop}%
\bibitem [{\citenamefont {Schwarz}(1988)}]{Schwarz1988}%
  \BibitemOpen
  \bibfield  {author} {\bibinfo {author} {\bibfnamefont {K.}~\bibnamefont
  {Schwarz}},\ }\href@noop {} {\bibfield  {journal} {\bibinfo  {journal}
  {Physical Review B}\ }\textbf {\bibinfo {volume} {38}},\ \bibinfo {pages}
  {2398} (\bibinfo {year} {1988})}\BibitemShut {NoStop}%
\bibitem [{\citenamefont {Kivotides}(2007)}]{Kivotides2007}%
  \BibitemOpen
  \bibfield  {author} {\bibinfo {author} {\bibfnamefont {D.}~\bibnamefont
  {Kivotides}},\ }\href@noop {} {\bibfield  {journal} {\bibinfo  {journal}
  {Physical Review B}\ }\textbf {\bibinfo {volume} {76}},\ \bibinfo {pages}
  {054503} (\bibinfo {year} {2007})}\BibitemShut {NoStop}%
\bibitem [{\citenamefont {Khomenko}\ \emph {et~al.}(2017)\citenamefont
  {Khomenko}, \citenamefont {Mishra},\ and\ \citenamefont
  {Pomyalov}}]{KhomenkoMishraPomyalov2017}%
  \BibitemOpen
  \bibfield  {author} {\bibinfo {author} {\bibfnamefont {D.}~\bibnamefont
  {Khomenko}}, \bibinfo {author} {\bibfnamefont {P.}~\bibnamefont {Mishra}},\
  and\ \bibinfo {author} {\bibfnamefont {A.}~\bibnamefont {Pomyalov}},\
  }\href@noop {} {\bibfield  {journal} {\bibinfo  {journal} {J. Low Temp. Phys.}\
  }\textbf {\bibinfo {volume} {13}},\ \bibinfo {pages} {1431} (\bibinfo {year}
  {2017})}\BibitemShut {NoStop}%
\bibitem [{\citenamefont {Yui}\ \emph {et~al.}(2018)\citenamefont {Yui},
  \citenamefont {Tsubota},\ and\ \citenamefont
  {Kobayashi}}]{YuiTsubotaKobayashi2018}%
  \BibitemOpen
  \bibfield  {author} {\bibinfo {author} {\bibfnamefont {S.}~\bibnamefont
  {Yui}}, \bibinfo {author} {\bibfnamefont {M.}~\bibnamefont {Tsubota}},\ and\
  \bibinfo {author} {\bibfnamefont {H.}~\bibnamefont {Kobayashi}},\ }\href@noop
  {} {\bibfield  {journal} {\bibinfo  {journal} {Phys. Rev. Lett.}\
  }\textbf {\bibinfo {volume} {120}},\ \bibinfo {pages} {155301} (\bibinfo
  {year} {2018})}\BibitemShut {NoStop}%
\bibitem [{\citenamefont {Choi}\ \emph {et~al.}(1998)\citenamefont {Choi},
  \citenamefont {Morgan},\ and\ \citenamefont
  {Burnett}}]{ChoiMorganBurnett1998}%
  \BibitemOpen
  \bibfield  {author} {\bibinfo {author} {\bibfnamefont {S.}~\bibnamefont
  {Choi}}, \bibinfo {author} {\bibfnamefont {A.}~\bibnamefont {Morgan}},\ and\
  \bibinfo {author} {\bibfnamefont {K.}~\bibnamefont {Burnett}},\ }\href@noop
  {} {\bibfield  {journal} {\bibinfo  {journal} {Phys. Rev. A}\ }\textbf
  {\bibinfo {volume} {57}},\ \bibinfo {pages} {4057} (\bibinfo {year}
  {1998})}\BibitemShut {NoStop}%
\bibitem [{\citenamefont {Kobayashi}\ and\ \citenamefont
  {Tsubota}(2006)}]{KobayashiTsubota2006}%
  \BibitemOpen
  \bibfield  {author} {\bibinfo {author} {\bibfnamefont {M.}~\bibnamefont
  {Kobayashi}}\ and\ \bibinfo {author} {\bibfnamefont {M.}~\bibnamefont
  {Tsubota}},\ }\href@noop {} {\bibfield  {journal} {\bibinfo  {journal}
  {Phys. Rev. Lett.}\ }\textbf {\bibinfo {volume} {97}},\ \bibinfo
  {pages} {145301} (\bibinfo {year} {2006})}\BibitemShut {NoStop}%
\bibitem [{\citenamefont {Madarassy}\ and\ \citenamefont
  {Barenghi}(2008)}]{MadarassyBarenghi2008}%
  \BibitemOpen
  \bibfield  {author} {\bibinfo {author} {\bibfnamefont {E.~J.~M.}\
  \bibnamefont {Madarassy}}\ and\ \bibinfo {author} {\bibfnamefont {C.~F.}\
  \bibnamefont {Barenghi}},\ }\href@noop {} {\bibfield  {journal} {\bibinfo
  {journal} {J. Low Temp. Phys.}\ }\textbf {\bibinfo {volume}
  {152}},\ \bibinfo {pages} {122} (\bibinfo {year} {2008})}\BibitemShut
  {NoStop}%
\bibitem [{\citenamefont {Wittmer}\ \emph {et~al.}(2021)\citenamefont
  {Wittmer}, \citenamefont {Schmied}, \citenamefont {Gasenzer},\ and\
  \citenamefont {Ewerz}}]{Wittmeretal2021}%
  \BibitemOpen
  \bibfield  {author} {\bibinfo {author} {\bibfnamefont {P.}~\bibnamefont
  {Wittmer}}, \bibinfo {author} {\bibfnamefont {C.-M.}\ \bibnamefont
  {Schmied}}, \bibinfo {author} {\bibfnamefont {T.}~\bibnamefont {Gasenzer}},\
  and\ \bibinfo {author} {\bibfnamefont {C.}~\bibnamefont {Ewerz}},\
  }\href@noop {} {\bibfield  {journal} {\bibinfo  {journal} {Phys. Rev.
  Lett.}\ }\textbf {\bibinfo {volume} {127}},\ \bibinfo {pages} {101601}
  (\bibinfo {year} {2021})}\BibitemShut {NoStop}%
\bibitem [{\citenamefont {Kobayashi}\ and\ \citenamefont
  {Tsubota}(2005)}]{KobayashiTsubota2005JPSJ}%
  \BibitemOpen
  \bibfield  {author} {\bibinfo {author} {\bibfnamefont {M.}~\bibnamefont
  {Kobayashi}}\ and\ \bibinfo {author} {\bibfnamefont {M.}~\bibnamefont
  {Tsubota}},\ }\href@noop {} {\bibfield  {journal} {\bibinfo  {journal} {J.
  Phys. Soc. Jpn.}\ }\textbf {\bibinfo {volume} {74}},\ \bibinfo {pages} {3248
  } (\bibinfo {year} {2005})}\BibitemShut {NoStop}%
\bibitem [{\citenamefont {Vinen}(1957)}]{Vinen1957}%
  \BibitemOpen
  \bibfield  {author} {\bibinfo {author} {\bibfnamefont {W.}~\bibnamefont
  {Vinen}},\ }\href@noop {} {\bibfield  {journal} {\bibinfo  {journal}
  {Proc. R. Soc. London. A}\ }\textbf {\bibinfo {volume} {242}},\ \bibinfo {pages} {493}
  (\bibinfo {year} {1957})}\BibitemShut {NoStop}%
\bibitem [{\citenamefont {Swanson}\ and\ \citenamefont
  {Donnelly}(1985)}]{SwansonDonnelly1985}%
  \BibitemOpen
  \bibfield  {author} {\bibinfo {author} {\bibfnamefont {C.~E.}\ \bibnamefont
  {Swanson}}\ and\ \bibinfo {author} {\bibfnamefont {R.~J.}\ \bibnamefont
  {Donnelly}},\ }\href@noop {} {\bibfield  {journal} {\bibinfo  {journal}
  {J. Low Temp. Phys.}\ }\textbf {\bibinfo {volume} {61}},\
  \bibinfo {pages} {363} (\bibinfo {year} {1985})}\BibitemShut {NoStop}%
\bibitem [{\citenamefont {Yoshida}\ \emph {et~al.}(2019)\citenamefont
  {Yoshida}, \citenamefont {Miura},\ and\ \citenamefont
  {Tsuji}}]{YoshidaMiuraTsuji2019}%
  \BibitemOpen
  \bibfield  {author} {\bibinfo {author} {\bibfnamefont {K.}~\bibnamefont
  {Yoshida}}, \bibinfo {author} {\bibfnamefont {H.}~\bibnamefont {Miura}},\
  and\ \bibinfo {author} {\bibfnamefont {Y.}~\bibnamefont {Tsuji}},\
  }\href@noop {} {\bibfield  {journal} {\bibinfo  {journal} {J. Low
  Temp. Phys.}\ }\textbf {\bibinfo {volume} {196}},\ \bibinfo {pages}
  {211 } (\bibinfo {year} {2019})}\BibitemShut {NoStop}%
\bibitem [{\citenamefont {Yoshida}\ \emph {et~al.}(2023)\citenamefont
  {Yoshida}, \citenamefont {Miura},\ and\ \citenamefont
  {Tsuji}}]{YoshidaMiuraTsuji2023}%
  \BibitemOpen
  \bibfield  {author} {\bibinfo {author} {\bibfnamefont {K.}~\bibnamefont
  {Yoshida}}, \bibinfo {author} {\bibfnamefont {H.}~\bibnamefont {Miura}},\
  and\ \bibinfo {author} {\bibfnamefont {Y.}~\bibnamefont {Tsuji}},\
  }\href@noop {} {\bibfield  {journal} {\bibinfo  {journal} {J. Low
  Temp. Phys.}\ }\textbf {\bibinfo {volume} {210}},\ \bibinfo {pages}
  {103 } (\bibinfo {year} {2023})}\BibitemShut {NoStop}%
\bibitem [{\citenamefont {Ohno}\ and\ \citenamefont
  {Ohtani}(2014)}]{OhnoOhtani2014}%
  \BibitemOpen
  \bibfield  {author} {\bibinfo {author} {\bibfnamefont {N.}~\bibnamefont
  {Ohno}}\ and\ \bibinfo {author} {\bibfnamefont {H.}~\bibnamefont {Ohtani}},\
  }\href@noop {} {\bibfield  {journal} {\bibinfo  {journal} {Plasma Fusion
  Res.}\ }\textbf {\bibinfo {volume} {9}},\ \bibinfo {pages} {3401071}
  (\bibinfo {year} {2014})}\BibitemShut {NoStop}%
\bibitem [{\citenamefont {Donnelly}\ and\ \citenamefont
  {Barenghi}(1998)}]{DonnellyBarenghi1998}%
  \BibitemOpen
  \bibfield  {author} {\bibinfo {author} {\bibfnamefont {R.~J.}\ \bibnamefont
  {Donnelly}}\ and\ \bibinfo {author} {\bibfnamefont {C.~F.}\ \bibnamefont
  {Barenghi}},\ }\href@noop {} {\bibfield  {journal} {\bibinfo  {journal}
  {J. Phys. Chem. Ref. Data}\ }\textbf {\bibinfo
  {volume} {27}},\ \bibinfo {pages} {1217} (\bibinfo {year}
  {1998})}\BibitemShut {NoStop}%
\bibitem [{\citenamefont {Tang}\ \emph {et~al.}(2021)\citenamefont {Tang},
  \citenamefont {Bao},\ and\ \citenamefont {Guo}}]{TangBaoGuo2021}%
  \BibitemOpen \bibfield  {author} {\bibinfo {author} {\bibfnamefont {Y.}~\bibnamefont
  {Tang}}, \bibinfo {author} {\bibfnamefont {S.}~\bibnamefont {Bao}},\ and\
  \bibinfo {author} {\bibfnamefont {W.}~\bibnamefont {Guo}},\ }\href@noop {}
  {\bibfield  {journal} {\bibinfo  {journal} {Proc. Natl. Acad. Sci. U.S.A.}\ }\textbf {\bibinfo
  {volume} {118}},\ \bibinfo {pages} {e2021957118} (\bibinfo {year}
    {2021})}\BibitemShut {NoStop}%
\bibitem [{\citenamefont {Chen}\ \emph {et~al.}(2022)\citenamefont {Chen},
  \citenamefont {Maruyama},\ and\ \citenamefont
  {Tsuji}}]{ChenMaruyamaTsuji2022}%
  \BibitemOpen
  \bibfield  {author} {\bibinfo {author} {\bibfnamefont {L.}~\bibnamefont
  {Chen}}, \bibinfo {author} {\bibfnamefont {T.}~\bibnamefont {Maruyama}},\
  and\ \bibinfo {author} {\bibfnamefont {Y.}~\bibnamefont {Tsuji}},\
  }\href@noop {} {\bibfield  {journal} {\bibinfo  {journal} {J. Low
  Temp. Phys.}\ }\textbf {\bibinfo {volume} {208}},\ \bibinfo {pages}
  {402} (\bibinfo {year} {2022})}\BibitemShut {NoStop}%
\bibitem [{\citenamefont {Brachet}\ \emph {et~al.}(2023)\citenamefont
  {Brachet}, \citenamefont {Sadaka}, \citenamefont {Zhang}, \citenamefont
  {Kalt},\ and\ \citenamefont {Danaila}}]{Brachetetal2023}%
  \BibitemOpen
  \bibfield  {author} {\bibinfo {author} {\bibfnamefont {M.}~\bibnamefont
  {Brachet}}, \bibinfo {author} {\bibfnamefont {G.}~\bibnamefont {Sadaka}},
  \bibinfo {author} {\bibfnamefont {Z.}~\bibnamefont {Zhang}}, \bibinfo
  {author} {\bibfnamefont {V.}~\bibnamefont {Kalt}},\ and\ \bibinfo {author}
  {\bibfnamefont {I.}~\bibnamefont {Danaila}},\ }\href@noop {} {\bibfield
  {journal} {\bibinfo  {journal} {J. Comp. Phys.}\ }\textbf
  {\bibinfo {volume} {488}},\ \bibinfo {pages} {112193} (\bibinfo {year}
  {2023})}\BibitemShut {NoStop}%
\bibitem [{\citenamefont {Griffin}\ \emph {et~al.}(2009)\citenamefont
  {Griffin}, \citenamefont {Nikuni},\ and\ \citenamefont
  {Zaremba}}]{GriffinNikuniZaremba2009}%
  \BibitemOpen
  \bibfield  {author} {\bibinfo {author} {\bibfnamefont {A.}~\bibnamefont
  {Griffin}}, \bibinfo {author} {\bibfnamefont {T.}~\bibnamefont {Nikuni}},\
  and\ \bibinfo {author} {\bibfnamefont {E.}~\bibnamefont {Zaremba}},\
  }\href@noop {} {\emph {\bibinfo {title} {Bose-Condensed Gases at Finite
  Temperatures}}}\ (\bibinfo  {publisher} {Cambridge University Press,
  Cambridge},\ \bibinfo {year} {2009})\BibitemShut {NoStop}%
\bibitem [{\citenamefont {Berloff}\ \emph {et~al.}(2014)\citenamefont
  {Berloff}, \citenamefont {Brachet},\ and\ \citenamefont
  {Proukakis}}]{BerloffBrachetProukakis2014}%
  \BibitemOpen
  \bibfield  {author} {\bibinfo {author} {\bibfnamefont {N.~G.}\ \bibnamefont
  {Berloff}}, \bibinfo {author} {\bibfnamefont {M.}~\bibnamefont {Brachet}},\
  and\ \bibinfo {author} {\bibfnamefont {N.~P.}\ \bibnamefont {Proukakis}},\
  }\href@noop {} {\bibfield  {journal} {\bibinfo  {journal} {Proc. Natl. Acad. Sci. U. S. A.}\ }\textbf
  {\bibinfo {volume} {111}},\ \bibinfo {pages} {4675} (\bibinfo {year}
  {2014})}\BibitemShut {NoStop}%
\end{thebibliography}

%
\end{document}